\documentclass[draft,journal,a4paper,oneside,onecolumn]{IEEEtran} 
\usepackage{amsmath,amsfonts, amssymb}

\newtheorem{thm}{Theorem}
\newtheorem{cor}[thm]{Corollary}
\newtheorem{lem}{Lemma}
\newtheorem{df}{Definition}
\newtheorem{rem}{Remark}

\newcommand{\GFq}{\mathrm{GF}(q)}
\newcommand{\A}{\mathcal{A}}
\newcommand{\B}{\mathcal{B}}
\newcommand{\C}{\mathcal{C}}
\newcommand{\cS}{\mathcal{S}}
\newcommand{\T}{\mathcal{T}}
\newcommand{\U}{\mathcal{U}}
\newcommand{\bU}{\overline{\mathcal{U}}}
\newcommand{\V}{\mathcal{V}}
\newcommand{\X}{\mathcal{X}}
\newcommand{\Y}{\mathcal{Y}}
\newcommand{\G}{\mathcal{G}}
\newcommand{\aalpha}{\boldsymbol{\alpha}}
\newcommand{\bbeta}{\boldsymbol{\beta}}
\newcommand{\kkappa}{\boldsymbol{\kappa}}
\newcommand{\cc}{\boldsymbol{c}}
\newcommand{\mm}{\boldsymbol{m}}
\newcommand{\uu}{\boldsymbol{u}}
\newcommand{\vv}{\boldsymbol{v}}
\newcommand{\ww}{\boldsymbol{w}}
\newcommand{\xx}{\boldsymbol{x}}
\newcommand{\yy}{\boldsymbol{y}}
\newcommand{\zz}{\boldsymbol{z}}
\newcommand{\lrB}[1]{\left[{#1}\right]}
\newcommand{\lrb}[1]{\left\{{#1}\right\}}
\newcommand{\lrsb}[1]{\left({#1}\right)}
\newcommand{\Error}{\mathrm{Error}}
\newcommand{\zero}{\boldsymbol{0}}
\newcommand{\one}{\boldsymbol{1}}
\newcommand{\limn}{\lim_{n\to\infty}}
\newcommand{\Encoder}{\varphi}
\newcommand{\Decoder}{\varphi^{-1}}
\newcommand{\im}{\mathrm{Im}}

\allowdisplaybreaks

\title{
Hash Property and Fixed-rate Universal Coding Theorems
}
\author{
Jun~Muramatsu~\IEEEmembership{Member,~IEEE,}
\and
Shigeki Miyake~\IEEEmembership{Member,~IEEE,}
 \thanks{J.~Muramatsu is with
        NTT Communication Science Laboratories, NTT Corporation,
        2-4, Hikaridai, Seika-cho, Soraku-gun, Kyoto 619-0237, Japan
        (E-mail: pure@cslab.kecl.ntt.co.jp).
        S.~Miyake is with
        NTT Network Innovation Laboratories, NTT Corporation,
        1-1, Hikarinooka, Yokosuka-shi, Kanagawa 239-0847, Japan
        (E-mail: miyake.shigeki@lab.ntt.co.jp).
  }
}

\begin{document}
\maketitle

\begin{abstract}
The aim of this paper is to prove the
achievability of fixed-rate universal coding problems
by using our previously introduced notion of hash property.
These problems are the fixed-rate lossless universal source coding
problem and the fixed-rate universal channel coding problem.
Since an ensemble of sparse matrices satisfies the hash property requirement,
it is proved that we can construct universal codes by using sparse matrices.
\end{abstract}

\section{Introduction}

The notion of hash property is introduced in \cite{HASH}.
It is a sufficient condition for the achievability of coding theorems
including lossless and lossy source coding, channel coding,
the Slepian-Wolf problem, the Wyner-Ziv problem, the Gel'fand-Pinsker problem,
and the problem of source coding with partial side information at the
decoder.
Since an ensemble of sparse matrices satisfies the hash property requirement,
it is proved that we can construct codes by using sparse matrices
and maximum-likelihood coding.

However, it is assumed in \cite{HASH} that
source and channel distributions are used when designing a code.
The aim of this paper is to prove fixed-rate universal coding theorems
based on the hash property,
where
a specific probability distribution is not assumed for the design
of a code
and the error probability of a code vanishes for all sources
specified by the encoding rate.

\begin{figure}[t]
\begin{center}
\unitlength 0.55mm
\begin{picture}(150,20)(0,0)
\put(4,7){\makebox(0,0){$M$}}
\put(10,7){\vector(1,0){10}}
\put(20,0){\framebox(14,14){$\Encoder$}}
\put(34,7){\vector(1,0){10}}
\put(50,7){\makebox(0,0){$X$}}
\put(56,7){\vector(1,0){10}}
\put(66,0){\framebox(24,14){$\mu_{Y|X}$}}
\put(90,7){\vector(1,0){10}}
\put(106,7){\makebox(0,0){$Y$}}
\put(112,7){\vector(1,0){10}}
\put(122,0){\framebox(14,14){$\Decoder$}}
\put(136,7){\vector(1,0){10}}
\put(150,7){\makebox(0,0){$M$}}
\put(75,-10){\makebox(0,0){$R<I(X;Y)$}}
\end{picture}
\end{center}
\caption{Channel Coding}
\label{fig:channel}
\end{figure}
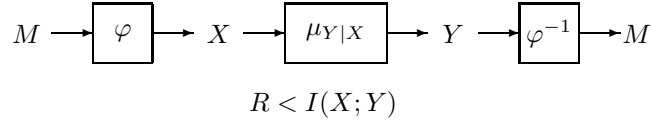

We prove theorems of fixed-rate lossless universal source coding
(see Fig.\ \ref{fig:source})
and fixed-rate universal channel coding (see Fig.\ \ref{fig:channel}).
In the construction of codes,
the maximum-likelihood coding used in \cite{HASH} is replaced
by a minimum-divergence encoder and a minimum-entropy decoder.
A practical algorithm has been obtained for
the minimum-entropy decoder by using linear programming \cite{CME05}.
It should be noted that
a practical algorithm for the minimum-divergence encoder
can also be obtained by using linear programming as shown in Section
\ref{sec:channelcoding}.
The fixed-rate lossless universal source coding theorem 
is proved in \cite{CSI82} for the ensemble of all linear matrices
in the context of the Slepian-Wolf source coding problem,
in \cite{K07} for the class of universal hash functions,
and in \cite{CRYPTLDPC} implicitly
for an ensemble of sparse matrices
in the context of a secret key agreement from correlated source
outputs.
The universal channel coding theorem that employs sparse matrices
is proved in \cite{Miyake07} for an additive noise channel
and in \cite{MM08} for an arbitrary channel.
It should be noted here that the linearity for an ensemble member
is not assumed in our proof.
Our proof assumes that
ensembles of sparse matrices have a hash property
and so is simpler
than previously reported proofs \cite{CRYPTLDPC}\cite{Miyake07}\cite{MM08}.

\begin{figure}[t]
\begin{center}
\unitlength 0.55mm
\begin{picture}(153,20)(0,18)
\put(5,25){\makebox(0,0){$X$}}
\put(10,25){\vector(1,0){10}}
\put(20,18){\framebox(14,14){$\Encoder$}}
\put(34,25){\vector(1,0){80}}
\put(74,29){\makebox(0,0){$R> H(X)$}}
\put(114,18){\framebox(14,14){$\Decoder$}}
\put(128,25){\vector(1,0){10}}
\put(148,25){\makebox(0,0){$X$}}
\end{picture}
\end{center}
\caption{Lossless Source Coding}
\label{fig:source}
\end{figure}
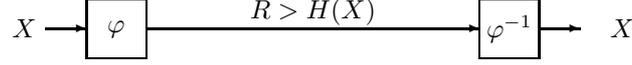

\section{Definitions and Notations}
Throughout this paper, we use the following definitions and notations.

Column vectors and sequences are denoted in boldface.
Let $A\uu$ denote a value taken by a function $A:\U^n\to\bU$ at $\uu\in\U^n$
where $\U^n$ is a domain of the function.
It should be noted that $A$ may be non-linear.
For a function $A$ and a set of functions $\A$, 
let $\im A$ and $\im \A$ be defined as
\begin{align*}
 \im A &\equiv \{A\uu: \uu\in\U^n\}
 \\
 \im\A &\equiv \bigcup_{A\in\A}\im A.
\end{align*}

The cardinality of a set $\U$ is denoted by $|\U|$
and $\U-\{\uu\}$ is a set difference.
We define sets $\C_A(\cc)$ and $\C_{AB}(\cc,\mm)$ as
\begin{align*}
 \C_A(\cc)
 &\equiv\{\uu: A\uu = \cc\}
 \\
 \C_{AB}(\cc,\mm)
 &\equiv\{\uu: A\uu = \cc, B\uu = \mm\}.
\end{align*}
In the context of linear codes, $\C_A(\cc)$ is called a coset determined
by $\cc$.

Let $p$ and $p'$ be probability distributions
and let $q$ and $q'$ be conditional probability distributions.
Then entropy $H(p)$, conditional entropy $H(q|p)$,
divergence $D(p\|p')$, and conditional divergence $D(q\|q'|p)$
are defined as
\begin{align*}
 H(p)
 &\equiv\sum_{u}p(u)\log\frac 1{p(u)}
 \\
 H(q|p)
 &\equiv\sum_{u,v}q(u|v)p(v)\log\frac 1{q(u|v)}
 \\
 D(p\parallel p')
 &\equiv
 \sum_{u}p(u)
 \log\frac{p(u)}{p'(u)}
 \\
 D(q\parallel q' | p)
 &\equiv
 \sum_{v} p(v)\sum_{u}q(u|v)
 \log\frac{q(u|v)}{q'(u|v)},
\end{align*}
where we assume the base $2$ of the logarithm.

Let $\mu_{UV}$ be the joint probability distribution of random variables
$U$ and $V$.
Let  $\mu_{U}$ and $\mu_{V}$ be the respective marginal distributions
and $\mu_{U|V}$ be the conditional probability distribution.
Then the entropy $H(U)$, the conditional entropy $H(U|V)$, and the mutual
information $I(U;V)$ of random variables are defined as
\begin{align*}
 H(U)&\equiv H(\mu_U)
 \\
 H(U|V)&\equiv H(\mu_{U|V}|\mu_{V})
 \\
 I(U;V)&\equiv H(\mu_U)+H(\mu_V)-H(\mu_{UV}).
\end{align*}

Let $\nu_{\uu}$ and $\nu_{\uu|\vv}$ be defined as
\begin{align*}
\nu_{\uu}(u)
 &\equiv
 \frac {|\{1\leq i\leq n : u_{i}=u\}|}n
 \\
 \nu_{\uu|\vv}(u|v)
 &\equiv \frac{\nu_{\uu\vv}(u,v)}{\nu_{\vv}(v)}.
\end{align*}
We call $\nu_{\uu}$ a type
\footnote{In~\cite{HASH},
the type of a sequence is defined as a histogram $\{n\nu_{\uu}(u)\}_{u\in\U}$.
}
of $\uu\in\U^n$
and $\nu_{\uu|\vv}$ a conditional type.
Let $U\equiv\nu_U$ be the type of a sequence
and $U|V\equiv\nu_{U|V}$ be the conditional type of a sequence
given a sequence of type $U$.
Then a set of typical sequences $\T_{U}$
and a set of conditionally typical sequences $\T_{U|V}(\vv)$
are defined as
\begin{align*}
 \T_{U}
 &\equiv
 \lrb{\uu:
 \nu_{\uu}=\nu_U
 }
 \\
 \T_{U|V}(\vv)
 &\equiv
 \lrb{\uu:
 \nu_{\uu|\vv}=\nu_{U|V}
 },
\end{align*}
respectively.
The empirical entropy, the empirical conditional entropy, and
empirical mutual information are defined as
\begin{align*}
 H(\uu)&\equiv H(\nu_{\uu})
 \\
 H(\uu|\vv)&\equiv H(\nu_{\uu|\vv}|\nu_{\vv})
 \\
 I(\uu;\vv)&\equiv H(\nu_{\uu})+H(\nu_{\vv})-H(\nu_{\uu\vv}).
\end{align*}

In the construction of a universal source code,
we use a {\em minimum-entropy decoder}
\[
 g_{A}(\cc)
 \equiv
 \arg\min_{\xx'\in\C_{A}(\cc)}H(\xx')
\]
It should be noted that the linear programing technique introduced in
\cite{CME05}
can be applied to the minimum-entropy decoder $g_A$.
In the construction of a universal channel code,
we use a {\em minimum-divergence encoder}
\[
 g_{AB}(\cc,\mm)
 \equiv\arg\min_{\xx'\in\C_{AB}(\cc,\mm)}D(\nu_{\xx'}\|\mu_X)
\]
and a minimum-entropy decoder
\[
 g_{A}(\cc,\yy)
 \equiv\arg\min_{\xx'\in\C_A(\cc)}H(\xx'|\yy).
\]
It should be noted that we have
\begin{align*}
 g_{AB}(\cc,\mm)
 &=\arg\max_{\xx'\in\C_{AB}(\cc,\mm)}\lrB{\log\mu_X(\xx')+nH(\nu_{\xx'})}
 \\
 &=\arg\max_{U'}\lrB{nH(U')
 +\max_{\xx'\in\C_{AB}(\cc,\mm)\cap\T_{U'}}\log\mu_X(\xx')}
\end{align*}
from Lemma~\ref{lem:exprob}.
When functions $A$ and $B$ are linear,
the linear programing technieque introduced in \cite{FWK05}
can be applied to the maximization $\max_{\xx'}\mu_X(\xx')$ because
$U'$ is fixed and the constraint condition
$\xx'\in\C_{AB}(\cc,\mm)\cap\T_{U'}$ is represented by linear functions.

Finally, we define $\chi(\cdot)$ as
\begin{align*}
 \chi(a = b)
 &\equiv
 \begin{cases}
	1,&\text{if}\ a = b
	\\
	0,&\text{if}\ a\neq b
 \end{cases}
 \\
 \chi(a \neq b)
 &\equiv
 \begin{cases}
	1,&\text{if}\ a \neq b
	\\
	0,&\text{if}\ a = b.
 \end{cases}
\end{align*}
We define a sequence
$\{\lambda_{\U}(n)\}_{n=1}^{\infty}$ as
\begin{align}
 \lambda_{\U}(n)
 &\equiv \frac{|\U|\log[n+1]}n.
 \label{eq:lambda}
\end{align}
It should be noted here that
the product set $\U\times\V$ is denoted by $\U\V$
when it appears in the subscript of this function
and we omit argument $n$ of $\lambda_{\U}$ when $n$ is clear in the context.
We define $|\cdot|^+$ as
\begin{equation}
 |\theta|^+
	\equiv
	\begin{cases}
	 \theta,&\text{if}\ \theta>0,\\
	 0,&\text{if}\ \theta\leq 0.
	\end{cases}
	\label{eq:plus}
\end{equation}

\section{$(\aalpha,\bbeta)$-hash Property}

In this section, we reveiw the notion of the
$(\aalpha,\bbeta)$-hash property introduced in \cite{HASH}.
This is a sufficient condition
for coding theorems, where the linearity of functions is not assumed.
By using this notion,
we prove a fixed-rate universal source coding theorem
and a fixed-rate universal source coding theorem.

Throughout the paper,
$A\uu$ denotes a value taken by a function $A$ at $\uu\in\U^n$
where $\U^n$ is the domain of the function.
It should again be noted here that $A$ may be non-linear.
We define the $(\aalpha,\bbeta)$-hash property in the following.
\begin{df}
 Let $\A$ be a set of functions $A:\U^n\to\bU$
 and we assume that $\im A=\im\A$ for all $A\in\A$ and
 \begin{equation}
	\limn \frac{\log\frac{|\bU|}{|\im\A|}}n=0.
	 \tag{H1}
	 \label{eq:imA}
 \end{equation}
 Let $p_A$ be a probability distribution on $\A$.
 We call a pair $(\A,p_A)$ an {\em ensemble}.
 Then, $(\A,p_A)$ has an $(\aalpha,\bbeta)$-{\em hash property} if
 $\aalpha\equiv\{\alpha(n)\}_{n=1}^{\infty}$ and
 $\bbeta\equiv\{\beta(n)\}_{n=1}^{\infty}$
 satisfy
 \begin{align}
	&\limn \alpha(n)=1
	\tag{H2}
	\label{eq:alpha}
	\\
	&\limn \beta(n)=0
	\tag{H3}
	\label{eq:beta}
 \end{align}
 and
 \begin{equation}
	\sum_{\substack{
	 \uu\in\T
	 \\
	\uu'\in\T'
	 }}
	 p\lrsb{\lrb{A: A\uu = A\uu'}}
	 \leq
	 |\T\cap\T'|
	 +\frac{|\T||\T'|\alpha(n)}{|\im\A|}
	 +\min\{|\T|,|\T'|\}\beta(n)
	 \tag{H4}
	 \label{eq:hash}
 \end{equation}
 for any $\T,\T'\subset\U^n$.
 Throughout this paper,
 we omit argument $n$ of $\alpha$ and $\beta$
 when $n$ is fixed.
 \hfill\QED
\end{df}

In the following, we present two examples of ensembles that have a hash
property.

\noindent{\bf Example 1:}
In this example, we consider
a universal class of hash functions introduced in \cite{CW}.
A set $\A$ of functions $A:\U^n\to\bU$ is called
a {\em universal class of hash functions} if 
\[
 |\lrb{A: A\uu=A\uu'}|\leq \frac{|\A|}{|\bU|}
\]
for any $\uu\neq\uu'$.
For example, the set of all functions on $\U^n$
and the set of all linear functions $A:\U^n\to\U^{l_A}$
are universal classes of hash functions (see \cite{CW}).

It should be noted that every example above
satisfies $\im\A=\bU$.
When $\A$ is a universal class of hash functions
and  $p_A$ is the uniform probability on $\A$,
we have
\begin{align*}
 \sum_{\substack{
 \uu\in\T
 \\
 \uu'\in\T'
 }}
 p_A\lrsb{\lrb{A: A\uu=A\uu'}}
 \leq
 |\T\cap\T'|+\frac{|\T||\T'|}{|\im\A|}.
\end{align*}
This implies that $(\A,p_A)$ has a $(\one,\zero)$-hash property,
where $\alpha(n)\equiv 1$ and $\beta(n)\equiv 0$ for every $n$.
\hfill\QED

\noindent{\bf Example 2:}
In this example,
we revew
the ensemble of $q$-ary sparse matrices introduced in \cite{HASH}.
In the following, let $\U\equiv\GFq$ and $l_A\equiv nR$.
We generate an $l\times n$ matrix $A$ with the following procedure:
\begin{enumerate}
 \item Start from an all-zero matrix.
 \item For each $i\in\{1,\ldots,n\}$, repeat the following
			 procedure $\tau$ times:
			 \begin{enumerate}
				\item Choose $(j,a)\in\{1,\ldots,l_A\}\times[\GFq-\{0\}]$
							uniformly at random.
				\item Add  $a$ to the  $(j,i)$ component of $A$.
			 \end{enumerate}
\end{enumerate}
Let $(\A,p_A)$ be an ensemble corresponding to the above procedure.
Then
\begin{align*}
 \im A
 &=
 \begin{cases}
	\lrb{
	\uu\in\U^l:
	\begin{aligned}
	 &\uu\ \text{has an even number of}
	 \\
	 &\text{non-zero elements}
	\end{aligned}
	 },
	 &\text{if}\ q=2
	 \\
	 \U^l,&\text{if}\ q>2
 \end{cases}
\end{align*}
for all $A\in\A$
and there is $(\aalpha_A,\bbeta_A)$ such that
$(\A,p_A)$ has an $(\aalpha_A,\bbeta_A)$-hash property (see \cite[Theorem 2]{HASH}).
\hfill\QED

In the following,
Let $\A$ (resp. $\B$) be a set of functions $A:\U^n\to\bU_A$
(resp. $B:\U^n\to\bU_B$).
We assume that
an ensemble $(\A,p_A)$
has an $(\aalpha_A,\bbeta_A)$-hash
property and an ensemble $(\A\times\B,p_{A}\times p_{B})$
also has an $(\aalpha_{AB},\bbeta_{AB})$-hash property.
We also assume that
$p_C$ and $p_M$ is the uniform distribution on $\im\A$ and $\im\B$,
respectively, and random variables $A$, $B$, $C$, and $M$ are mutually
independent, that is,
\begin{align*}
 p_C(\cc)&=
 \begin{cases}
	\frac 1{|\im\A|},&\text{if}\ \cc\in\im\A
	\\
	0,&\text{if}\ \cc\in\bU-\im\A
 \end{cases}
 \\
 p_M(\mm)&=
 \begin{cases}
	\frac 1{|\im\B|},&\text{if}\ \mm\in\im\B
	\\
	0,&\text{if}\ \mm\in\bU-\im\A
 \end{cases}
 \\
 p_{ABCM}(A,B,\cc,\mm)&=p_{A}(A)p_B(B)p_C(\cc)p_M(\mm)
\end{align*}
for any $A$, $B$, and $\cc$.
We use the following lemmas, which are shown in \cite{HASH}.
\begin{lem}[{\cite[Lemma 9]{HASH}}]
 \label{lem:E}
 For any $A$ and $\uu\in\U^n$,
 \begin{equation*}
	 p_C\lrsb{\lrb{\cc: A\uu=\cc}}
	 =
	 \sum_{c}p_C(\cc)\chi(A\uu=\cc)
	 =\frac 1{|\im\A|}
 \end{equation*}
 and for any $\uu\in\U^n$,
 \begin{equation*}
	 E_{AC}\lrB{\chi(A\uu=\cc)}
	 =\sum_{A,\cc}p_{AC}(A,\cc)\chi(A\uu=\cc)
	 =\frac 1{|\im\A|}.
 \end{equation*}
\end{lem}
\begin{lem}[{\cite[Lemma 2]{HASH}}]
 \label{lem:Anotempty}
 If $\G\subset\U^n$ and $\uu\notin\G$, then
\[
 p_A\lrsb{\lrb{
 A: \G\cap\C_A(A\uu)\neq \emptyset
 }}
 \leq 
 \frac{|\G|\alpha_A}{|\im\A|} + \beta_A.
\]
\hfill\QED
\end{lem}

\begin{lem}[{\cite[Lemma 5]{HASH}}]
 \label{lem:noempty}
 If $\T\neq\emptyset$, then
 \begin{equation*}
	p_{ABCM}\lrsb{\lrb{(A,B,\cc,\mm):
	\T\cap\C_{AB}(\cc,\mm)=\emptyset
	}}
	\leq
 \alpha_{AB}-1+\frac{|\im\A||\im\B|\lrB{\beta_{AB}+1}}{|\T|}.
 \end{equation*}
 \hfill\QED
\end{lem}

When $(\A,p_A)$ and $(\B,p_B)$
are the ensembles of $l_A\times n$ and $l_B\times n$ linear matrices,
respectively,
we have the following lemma.
\begin{lem}[{\cite[Lemma 7]{HASH}}]
 \label{lem:hash-linApB}
 The joint distribution $(\A\times\B,p_{AB})$ 
 has an $(\aalpha_{AB},\bbeta_{AB})$-hash property
 for the ensemble of functions $A\oplus B:\U^n\to\U^{l_A+l_B}$ defined as
 \[
 A\oplus B(\uu)\equiv(A\uu,B\uu),
 \]
 where
 \begin{align}
	\alpha_{AB}(n)&=\alpha_A(n)\alpha_B(n)
	\label{eq:alphaAB}
	\\
	\beta_{AB}(n)&=\min\{\beta_A(n),\beta_B(n)\}.
	\label{eq:betaAB}
 \end{align}
 \hfill\QED
\end{lem}

\section{Fixed-rate Lossless Universal Source Coding}
In this section, we consider the fixed-rate lossless universal source coding
illustrated in Fig.\ \ref{fig:source}.

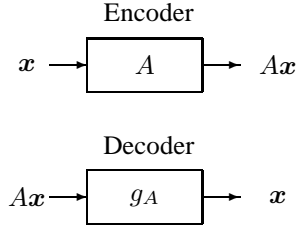
\begin{figure}[t]
\begin{center}
\unitlength 0.5mm
\begin{picture}(176,25)(0,30)
\put(82,41){\makebox(0,0){Encoder}}
\put(50,27){\makebox(0,0){$\xx$}}
\put(56,27){\vector(1,0){10}}
\put(66,20){\framebox(30,14){$A$}}
\put(96,27){\vector(1,0){10}}
\put(116,27){\makebox(0,0){$A\xx$}}
\end{picture}
\\[5mm]
\begin{picture}(176,25)(0,30)
\put(82,41){\makebox(0,0){Decoder}}
\put(50,27){\makebox(0,0){$A\xx$}}
\put(56,27){\vector(1,0){10}}
 \put(66,20){\framebox(30,14){$g_{A}$}}
\put(96,27){\vector(1,0){10}}
\put(116,27){\makebox(0,0){$\xx$}}
\end{picture}
\end{center}
\caption{Construction of Fixed-rate Source Code}
\label{fig:source-code}
\end{figure}

For a given encoding rate $R$,
$l_A$ is given by
\begin{align*}
 l_A\equiv\frac{nR}{\log|\X|}.
\end{align*}
We fix a function
\begin{align*}
 A&:\X^n\to\X^{l_A}
\end{align*}
which is available to construct an encoder and a decoder.
We define the encoder and the decoder (illustrated in Fig.\ \ref{fig:source-code})
\begin{align*}
 \Encoder_X&:\X^n\to\X^{l_A}
 \\
 \Decoder&:\X^{l_{A}}\to\X^n
\end{align*}
as
\begin{align*}
 \Encoder(\xx)&\equiv A\xx
 \\
 \Decoder(\cc)
 &\equiv g_{A}(\cc),
\end{align*}
where
\begin{align*}
 g_{A}(\cc)
 &\equiv
 \arg\min_{\xx'\in\C_{A}(\cc)}H(\xx').
\end{align*}

The error probability $\Error_X(A)$ is given by
\begin{align*}
 \Error_X(A)
 &
 \equiv
 \mu_{X}\lrsb{\lrb{
	\xx:
	\Decoder(\Encoder(\xx))
	\neq \xx
 }}.
\end{align*}

We have the following theorem.
It should be noted that
the alphabet $\X$ may not be binary.
\begin{thm}
 \label{thm:source}
 Assume that an ensemble $(\A,p_A)$ has an $(\aalpha_A,\bbeta_A)$-hash property.
 For a fixed rate $R$, $\delta>0$
 and a sufficiently large $n$,
 there is a function (matrix) $A\in\A$
 such that
 \begin{equation}
	\Error_X(A)\leq
	 \max\lrb{\frac{\alpha_A|\X|^{l_A}}{|\im\A|},1}
	 2^{-n[\inf F_X(R)-2\lambda_{\X}]}
	 +\beta_{A}
	 \label{eq:error-source}
 \end{equation}
 for any stationary memoryless sources $X$ satisfying
 \begin{align}
	H(X)<R,
	\label{eq:rate-source}
 \end{align}
 where
 \begin{align*}
	F_X(R)
	&\equiv\min_{U'}\lrB{D(\nu_{U'}\|\mu_X)+|R-H(U')|^+}
 \end{align*}
 and the infimum is taken over all $X$ satisfying (\ref{eq:rate-source}).
 Since
 \[
 \inf_{X:H(X)>R}F_X(R)>0,
 \]
 then the error probability goes to zero as
 $n\to\infty$ for all $X$ satisfying (\ref{eq:rate-source}).
 \hfill\QED
\end{thm}

We can prove the coding theorem for 
a channel $\mu_{Y|X}$ with additive noise $Z\equiv Y-X$
by letting $A$ and $\C_A(\zero)=\{\xx: A\xx=\zero\}$ 
be a parity check matrix and a set of codewords
(channel inputs), respectively.
Then the encoding rate of this channel code is given by
\[
 \frac{\log|\C_A(\zero)|}n \geq \log|\X|-R
\]
and the error probability is given as
\begin{align*}
 \Error_{Y|X}(A)
 \equiv
 \frac 1{|\C_A(\zero)|}
 \sum_{\xx\in\C_A(\zero)}
 \mu_{Y|X}\lrsb{
 \lrb{\yy: g_A(A\yy)\neq \yy-\xx}\left|\right.\xx
 }.
\end{align*}
Since
\begin{align*}
 \zz &= \yy - \xx
 \\
 A\zz &= A\yy-A\xx = A\yy,
\end{align*}
then  the decoding of channel input $\xx$ from a syndrome $A\yy$
is equivalent to
the decoding of source output $\zz$ from its codeword $A\zz$
by using $g_A$.
We have the following corollary.
\begin{cor}
 \label{thm:additive-channel}
 Assume that an ensemble $(\A,p_A)$ of linear functions
 has an $(\aalpha_A,\bbeta_A)$-hash property.
 For a fixed rate $R$,  $\delta>0$
 and sufficiently large $n$,
 there is a (sparse) matrix $A\in\A$
 such that
 \begin{equation*}
	\Error_{Y|X}(A)\leq
	 \max\lrb{\frac{\alpha_A|\X|^{l_A}}{|\im\A|},1}
	 2^{-n[\inf F_Z(R)-2\lambda_{\X}]}
	 +\beta_{A}
	 \label{eq:error-additive-noise}
 \end{equation*}
 for any stationary memoryless channel with additive noize $Z$ satisfying
 \begin{align}
	\log|\X|-R<I(X;Y)=\log|\X|-H(Z),
	\label{eq:rate-additive-noise}
 \end{align}
 where the infimum is taken over all $Z$ satisfying
 (\ref{eq:rate-additive-noise}) and
 the error probability goes to zero as
 $n\to\infty$ for all $X$ satisfying (\ref{eq:rate-additive-noise}).
 \hfill\QED
\end{cor}

\begin{rem}
 It should be noted here that
 the condition (\ref{eq:alpha}) can be replaced by
 \begin{equation}
	\limn\frac{\log\alpha_A(n)}n=0.
	 \label{eq:logalpha}
 \end{equation}
 By using the expurgation technique described in \cite{BB04},
 we obtain an ensemble of sparce matrices
 that have an $(\aalpha_A,\zero)$-hash property,
 where (\ref{eq:alpha}) is replaced by (\ref{eq:logalpha}).
 This implies that we can omit the term $\beta_A$
 from the upper bound of the error probability.
 \hfill\QED
\end{rem}

\begin{rem}
 Since a class of universal hash functions
 with a uniform distribution
 and an ensemble of all linear functions
 has a $(\one,\zero)$-hash property,
 we obtain the same results as those reported in \cite{K07} and \cite{CSI82},
 respectively, where $F_X$ represents the error exponent function.
 When $(\A,p_A)$ is an ensemble of sparse matrices
 and $(\aalpha_A,\bbeta_A)$ is defined properly,
 we have the same result as that found in \cite{Miyake07}.
 \hfill\QED
\end{rem}

\section{Fixed-rate Universal Channel Coding}
\label{sec:channelcoding}

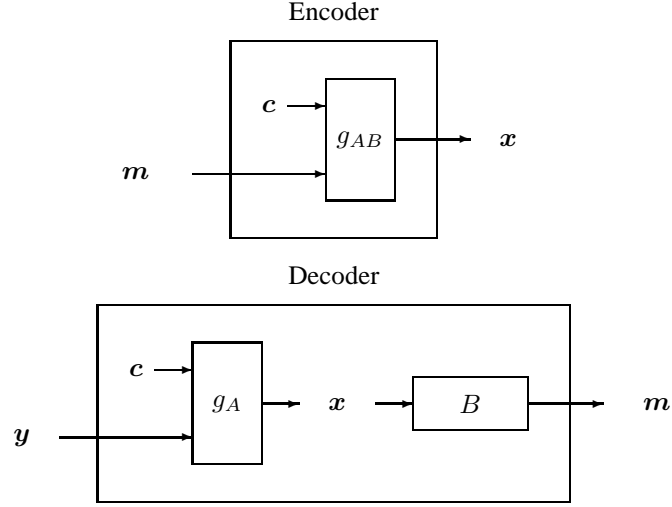
\begin{figure}[t]
\begin{center}
\unitlength 0.5mm
\begin{picture}(176,70)(0,0)
\put(82,60){\makebox(0,0){Encoder}}
\put(65,35){\makebox(0,0){$\cc$}}
\put(70,35){\vector(1,0){10}}
\put(30,17){\makebox(0,0){$\mm$}}
\put(45,17){\vector(1,0){35}}
\put(80,10){\framebox(18,32){$g_{AB}$}}
\put(98,26){\vector(1,0){20}}
\put(128,26){\makebox(0,0){$\xx$}}
\put(55,0){\framebox(54,52){}}
\end{picture}
\\
\begin{picture}(176,70)(0,0)
\put(82,60){\makebox(0,0){Decoder}}
\put(30,35){\makebox(0,0){$\cc$}}
\put(35,35){\vector(1,0){10}}
\put(0,17){\makebox(0,0){$\yy$}}
\put(10,17){\vector(1,0){35}}
\put(45,10){\framebox(18,32){$g_A$}}
\put(63,26){\vector(1,0){10}}
\put(83,26){\makebox(0,0){$\xx$}}
\put(93,26){\vector(1,0){10}}
\put(103,19){\framebox(30,14){$B$}}
\put(133,26){\vector(1,0){20}}
\put(167,26){\makebox(0,0){$\mm$}}
\put(20,0){\framebox(124,52){}}
\end{picture}
\end{center}
\caption{Construction of Channel Code}
\label{fig:channel-code}
\end{figure}

The code for the channel coding problem
(illustrated in Fig.\ \ref{fig:channel})
is given in the following (illustrated in Fig.\ \ref{fig:channel-code}).
The idea for the construction is drawn from \cite{SWLDPC}\cite{HASH}\cite{MM08}.
We give the explicit construction of the encoder by using
minimum-divergence encoding,
which is not described 
in \cite{SWLDPC}\cite{HASH}\cite{MM08}.

For a given $R_A,R_B>0$,
let
\begin{align*}
 A&:\X^n\to\X^{l_A}
 \\
 B&:\X^n\to\X^{l_B}
\end{align*}
satisfying
\begin{align*}
 R_A&=\frac{\log|\im A|}n
 \\
 R_B&=\frac{\log|\im B|}n,
\end{align*}
respectively.

We fix functions $A$, $B$ and a vector $\cc_n\in\X^{l_A}$
available to constract an encoder and a decoder.

We define the encoder and the decoder
\begin{align*}
 \Encoder&:\X^{l_B}\to\X^n
 \\
 \Decoder&:\Y^n\to\X^{l_B}
\end{align*}
as
\begin{align*}
 \Encoder(\mm)
 &\equiv g_{AB}(\cc,\mm)
 \\
 \Decoder(\yy)
 &\equiv Bg_A(\cc,\yy),
\end{align*}
where
\begin{align*}
 g_{AB}(\cc,\mm)
 &\equiv\arg\min_{\xx'\in\C_{AB}(\cc,\mm)}D(\nu_{\xx'}\|\mu_X)
 \\
 g_{A}(\cc,\yy)
 &\equiv\arg\min_{\xx'\in\C_A(\cc)}H(\xx'|\yy).
\end{align*}

The error probability $\Error_{Y|X}(A,B,\cc)$ is given by
\begin{align*}
 \Error_{Y|X}(A,B,\cc)
 \equiv
 \sum_{\mm,\yy}p_M(\mm)\mu_{Y|X}(\yy|\Encoder(\mm))
 \chi(\Decoder(\yy)\neq\mm),
\end{align*}
where
\begin{align*}
 p_M(\mm)
 &\equiv
 \begin{cases}
	\frac 1{|\im B|},
	&\quad\text{if}\ \cc\in\im B
	\\
	0
	&\quad\text{if}\ \cc\notin\im B.
 \end{cases}
\end{align*}
It should be noted that $\im B$ represents a set of all messages
and $R_B$ represents the encoding rate of a channel.

We have the following theorem.
\begin{thm}
 \label{thm:channel}
 Assume that an ensemble
 $(\A,p_A)$ (resp.\ $(\A\times\B,p_{AB})$)
 has an $(\aalpha_A,\bbeta_A)$-hash (resp.\ $(\aalpha_{AB},\bbeta_{AB})$-hash)
 property.
 For a fixed rate $R_A,R_B>0$, a given input distribution $\mu_X$
 satisfying
 \begin{align}
	H(X)&>R_A+R_B,
	 \label{eq:rateAB}
 \end{align}
 $\delta>0$, and a sufficiently large $n$,
 there are functions (matrices) $A\in\A$, $B\in\B$, and a vector
 $\cc\in\im A$
 such that
 \begin{equation}
	\Error_{Y|X}(A,B,\cc)
	 \leq
	 \alpha_{AB}-1
	 +
	 \frac{\beta_{AB}+1}{\kappa}
	 +
	 2\kappa
	 \lrB{
	 \max\lrb{\alpha_A,1}
	 2^{-n[\inf F_{Y|X}(R_A)-2\lambda_{\X\Y}]}
	 +
	 \beta_A
	 }
	 \label{eq:error-channel}
 \end{equation}
 for all $\mu_{Y|X}$ satisfying
 \begin{align}
	H(X|Y)&<R_A,
	\label{eq:rateA}
 \end{align}
 where
 \[
 F_{Y|X}(R)\equiv
 \min_{V|U}[D(\nu_{V|U}\|\mu_{Y|X}|\nu_U)+|R-H(U|V)|^+],
 \]
 the infimum is taken over all $\mu_{Y|X}$ satisfying (\ref{eq:rateAB}),
 and $\kkappa\equiv\{\kappa(n)\}_{n=1}^{\infty}$ is an arbitrary
 sequence satisfying
 \begin{gather}
	\limn\kappa(n)=\infty
	\label{eq:c-k1}
	\\
	\limn \kappa(n)\beta_A(n)=0
	\label{eq:c-k2}
	\\
	\limn\frac{\log\kappa(n)}n=0
	\label{eq:c-k3}
 \end{gather}
 and $\kappa$ denotes $\kappa(n)$.
 Since
 \[
 \inf_{\substack{
 \mu_{Y|X}:\\
 H(Y|X)<R_A
 }}
 F_{Y|X}(R_A)>0,
 \]
 then the right hand side of (\ref{eq:error-channel}) goes to zero as
 $n\to\infty$
 for all $\mu_{Y|X}$ satisfying (\ref{eq:rateA}).
 \hfill\QED
\end{thm}
\begin{rem}
 It should be noted here that we have
 \begin{align}
	I(X;Y)&>R_B
	\label{eq:rateB}
 \end{align}
 from (\ref{eq:rateA}) and (\ref{eq:rateAB}).
 However (\ref{eq:rateA}) and (\ref{eq:rateB}) do not imply
 (\ref{eq:rateAB})
 even when $R_A<H(X)$.
 \hfill\QED
\end{rem}
\begin{rem}
 For $\bbeta_A$ satisfying (\ref{eq:beta}),
 there is $\kkappa$ satisfying (\ref{eq:c-k1})--(\ref{eq:c-k3}) by letting
 \begin{equation}
	\kappa(n)\equiv
	 \begin{cases}
		n^{\xi}
		&\text{if}\ \beta_A(n)=o\lrsb{n^{-\xi}}
		\\
		\frac 1{\sqrt{\beta_A(n)}},
		&\text{otherwise}
	 \end{cases}
	 \label{eq:kappa}
 \end{equation}
 for every $n$.
 If $\beta_A(n)$ is not $o\lrsb{n^{-\xi}}$,
 there is $\kappa'>0$ such that
 $\beta_A(n)n^{\xi}>\kappa'$
 and
 \begin{align*}
	\frac{\log\kappa(n)}n
	&=
	\frac{\log\frac 1{\beta_A(n)}}{2n}
	\\
	&\leq \frac{\log\frac{n^{\xi}}{\kappa'}}{2n}
	\\
	&=\frac{\xi\log n-\log\kappa'}{2n}
 \end{align*}
 for all sufficiently large $n$.
 This implies that $\kkappa$ satisfies (\ref{eq:c-k3}).
 It should be noted that
 we can let $\xi$ be arbitrarily large in (\ref{eq:kappa})
 when $\beta_A(n)$ vanishes exponentially fast.
 This parameter $\xi$ affects the upper bound of (\ref{eq:error-channel}).
 \hfill\QED
\end{rem}

\begin{rem}
 From Lemma~\ref{lem:hash-linApB},
 we have the fact that
 the condition (\ref{eq:beta}) of $\bbeta_B$ is not necessary
 for the ensembles $(\A,p_A)$ and $(\B,p_B)$ of linear functions.
 \hfill\QED
\end{rem}

\section{Proof of Theorems}
In this section, we prove the theorems.

\subsection{Proof of Theorem \ref{thm:source}}
Let
\begin{align*}
 \G_U&\equiv\lrb{
 \xx': H(\xx')\leq H(U)
 }.
\end{align*}
If $\xx\in\T_U$ and $g_A(A\xx)\neq \xx$,
then there is $\xx'\in\C_A(\A\xx)$ such that $\xx'\neq\xx$
and
\[
 H(\xx')\leq H(\xx)=H(U),
\]
which implies that
\begin{align*}
 &\lrB{\G_U-\{\xx\}}\cap\C_{A}(A\xx)\neq\emptyset.
\end{align*}
Then we have
\begin{align}
 E_{A}\lrB{\Error_X(A)}
 &=
 E_{A}\lrB{\sum_{\xx}\mu_X(\xx)\chi(g_A(A\xx)\neq\xx)}
 \notag
 \\
 &\leq
 \sum_U\sum_{\xx\in\T_U}\mu_{X}(\xx)
 p_{A}\lrsb{\lrb{
 \begin{aligned}
	&A:\\
	&\lrB{\G_U-\{\xx\}}\cap\C_{A}(A\xx)\neq\emptyset
 \end{aligned} 
 }}
 \notag
 \\
 &\leq
 \sum_{U}\sum_{\xx\in\T_U}\mu_{X}(\xx)
 \max\lrb{
 \frac{|\G_U|\alpha_{A}}{|\im\A|}
 +\beta_{A},
 1
 }
 \notag
 \\
 &\leq
 \sum_{U}\sum_{\xx\in\T_U}\mu_{X}(\xx)
 \max\lrb{
 \frac{|\X|^{l_A}2^{-n[R-H(U)-\lambda_{\X}]}\alpha_{A}}{|\im\A|},
 1
 }
 \notag
 +\beta_{A}
 \notag
 \\
 &\leq
 \max\lrb{\frac{\alpha_A|\X|^{l_A}}{|\im\A|},1}
 \sum_U
 2^{-n[D(\nu_{U}\|\mu_X)+|R-H(U)|^+-\lambda_{\X}]}
 \notag
 +\beta_{A}
 \notag
 \\
 &\leq
 \max\lrb{\frac{\alpha_A|\X|^{l_A}}{|\im\A|},1}
 2^{-n[F_X(R)-2\lambda_{\X}]}
 +\beta_{A},
 \notag
\end{align}
where
the second inequality comes from Lemma~\ref{lem:Anotempty},
the third inequality comes from Lemma~\ref{lem:typical-number},
the fourth inequality comes from
Lemmas~\ref{lem:typenumber} and \ref{lem:exprob},
and the last inequality comes from the definition of $F_X$
and Lemma \ref{lem:typebound}.
Then we have the fact
that there is a function (matrix) $A\in\A$ satisfying (\ref{eq:error-source}).
\hfill\QED

\subsection{Proof of Theorem~\ref{thm:channel}}

Let $UV\equiv\nu_{VU}$ be a joint type of the sequence
$(\xx,\yy)\in\X^n\times\Y^n$, where
the marginal type $U$ is defined as
\begin{align}
 U\equiv\arg\min_{U'}D(\nu_{U'}\|\mu_{X}).
 \label{eq:U}
\end{align}
and the conditional type given type $U$ is denoted by $V|U$.
Since $R_A+R_B<H(X)$
and $H(U)$ approaches $H(X)$ as $n$ goes to infinity
because of the law of large numbers and the continuity of the entropy function,
we have
\[
H(U)-\lambda_{\X}>R_A+R_B+\frac{\log\kappa}n
\]
for all sufficiently large $n$.
Then we have
\begin{align*}
 |\T_{U}|
 &\geq 2^{n[H(U)-\lambda_{\X}]},
 \\
 &\geq
 \kappa
 2^{n[R_A+R_B]}
 \\
 &=
 \kappa
 |\im\A||\im\B|
\end{align*}
for all sufficiently large $n$,
where the first inequality comes from Lemma \ref{lem:typenumber}.
This implies that
there is $\T\subset\T_{U}$ such that
\begin{align}
 \kappa
 &\leq
 \frac{|\T|}{|\im\A||\im\B|}
 \leq 
 2\kappa
 \label{eq:TU}
\end{align}
for all sufficiently large $n$.

Let
\begin{align}
 &\bullet g_{AB}(\cc,\mm)\in\T
 \tag{UC1}
 \\
 &\bullet g_{A}(\cc,\yy)=g_{AB}(\cc,\mm).
 \tag{UC2}
\end{align}
Then we have
\begin{align}
 \Error(A,B,\cc,\mu_{Y|X})
 \leq
 p_{MY}(\cS_1^c)
 +
 p_{MY}(\cS_1\cap\cS_2^c),
 \label{eq:C0}
\end{align}
where
\begin{align*}
 \cS_i
 &\equiv
 \lrb{
 (\mm,\yy,\ww): \text{(UC$i$)}
 }.
\end{align*}

First, we evaluate $E_{ABC}\lrB{p_{MY}(\cS_1^c)}$.
We have
\begin{align}
 E_{ABC}\lrB{p_{MY}(\cS_1^c)}
 &=
 p_{ABCM}
 \lrsb{\lrb{
	(A,B,\cc,\mm):
	\T\cap\C_{AB}(\cc,\mm)=\emptyset
 }}
 \notag
 \\
 &\leq 
 \alpha_{AB}-1
 +
 \frac{|\im\A||\im\B|\lrB{\beta_{AB}+1}}{|\T|}
 \notag
 \\
 &\leq 
 \alpha_{AB}-1
 +
 \frac{\beta_{AB}+1}{\kappa}
 \label{eq:C1}
\end{align}
where the equality comes from the property of $\T$,
the first inequailty comes from Lemma \ref{lem:noempty}
and the second inequality comes from (\ref{eq:TU}).

Next, we evaluate $E_{ABC}\lrB{p_{MY}(\cS_1\cap\cS_2^c)}$.
Let
\[
 \G(\yy)\equiv\{\xx': H(\xx'|\yy)\leq H(U|V)\}
\]
and assume that $(\xx,\yy)\in\T_{UV}$.
Then we have
\begin{align}
 E_{AC}\lrB{
 \chi(A\xx=\cc)\chi(g_A(\cc,\yy)\neq \xx)
 }
 &=
 p_{AC}\lrsb{\lrb{
 (A,\cc):
 \begin{aligned}
	&A\xx=\cc
	\\
	&\exists\xx'\neq\xx\ \text{s.t.}
	\\
	&H(\xx'|\yy)\leq H(\xx|\yy)\ \text{and}\ A\xx'=\cc
 \end{aligned}
 }}
 \notag
 \\
 &=
 p_{A}\lrsb{\lrb{
 A:
 \begin{aligned}
	&\exists\xx'\neq\xx\ \text{s.t.}
	\\
	&H(\xx'|\yy)\leq H(\xx|\yy)\ \text{and}\ A\xx'=A\xx
 \end{aligned}
 }}
 \notag
 p_C\lrsb{\lrb{
 \cc: A\xx=\cc
 }}
 \notag
 \\
 &=
 \frac 1{|\im\A|}
 p_{A}\lrsb{\lrb{
 A:
 \begin{aligned}
	&\exists\xx'\neq\xx\ \text{s.t.}\ H(\xx'|\yy)\leq H(U|V)
	\\
	&\text{and}\ A\xx'=A\xx
 \end{aligned}
 }}
 \notag
 \\
 &\leq
 \frac 1{|\im\A|}
 \max\lrb{
 \sum_{\xx'\in[\G(\yy)-\{\xx\}]}
 p_A\lrsb{\lrb{A: A\xx=A\xx'}},
 1
 }
 \notag
 \\
 &\leq
 \frac 1{|\im\A|}
 \max\lrb{
 \frac{2^{n[H(U|V)+\lambda_{\X\Y}]}\alpha_A}
 {|\im\A|}
 +\beta_A,
 1
 }
 \notag
 \\
 &=
 \frac 1{|\im\A|}
 \max\lrb{
 2^{-n[R_A-H(U|V)-\lambda_{\X\Y}]}\alpha_A
 +\beta_A,
 1
 }
 \notag
 \\
 &\leq
 \frac 1{|\im\A|}
 \lrB{
 \max\lrb{\alpha_A,1}
 2^{-n[|R_A-H(U|V)|^+
 -\lambda_{\X\Y}]}
 +\beta_A
 },
 \label{eq:C2sub}
\end{align}
where $|\cdot|^+$ is defined by (\ref{eq:plus}),
the third equality comes from Lemma \ref{lem:E} and
the second inequality comes
from Lemma \ref{lem:typical-number}
and (\ref{eq:hash}) for an ensemble $p_A$.
Then we have
\begin{align}
 &
 E_{ABC}\lrB{p_{MY}(\cS_1\cap\cS_2^c)}
 \notag
 \\*
 &=
 E_{ABCM}\left[
 \sum_{\xx\in\T}\sum_{V|U}
 \sum_{\yy\in\T_{V|U}(\xx)}
 \mu_{Y|X}(\yy|\xx)
 \chi(g_{AB}(\cc,\mm)=\xx)\chi(g_A(\cc,\yy)\neq \xx)
 \right]
 \notag
 \\
 &\leq
 E_{ABCM}\left[
 \sum_{\xx\in\T}\sum_{V|U}
 \sum_{\yy\in\T_{V|U}(\xx)}
 \mu_{Y|X}(\yy|\xx)
 \chi(A\xx=\cc)\chi(B\xx=\mm)\chi(g_A(\cc,\yy)\neq \xx)
 \right]
 \notag
 \\
 &=
 \sum_{\xx\in\T}
 \sum_{V|U}\sum_{\yy\in\T_{V|U}(\xx)}
 \mu_{Y|X}(\yy|\xx)
 E_{AC}\left[
 \chi(A\xx=\cc)\chi(g_A(\cc,\yy)\neq \xx)
 \right]
 E_{BM}\lrB{
 \chi(B\xx=\mm)
 }
 \notag
 \\
 &
 \leq
 \frac 1{|\im\A||\im\B|}
 \sum_{\xx\in\T}
 \sum_{V|U}
 \sum_{\yy\in\T_{V|U}(\xx)}
 \mu_{Y|X}(\yy|\xx)
 \lrB{
 \max\lrb{\alpha_A,1}
 2^{-n[|R_A-H(U|V)|^+
 -\lambda_{\X\Y}]}
 +\beta_A
 }
 \notag
 \\
 &
 =
 \frac 1{|\im\A||\im\B|}
 \sum_{\xx\in\T}
 \lrB{
 \sum_{V|U}
 \sum_{\yy\in\T_{V|U}(\xx)}
 \mu_{Y|X}(\yy|\xx)
 \max\lrb{\alpha_A,1}
 2^{-n[|R_A-H(U|V)|^+
 -\lambda_{\X\Y}]}
 +
 \beta_A
 }
 \notag
 \\
 &
 \leq
 \frac 1{|\im\A||\im\B|}
 \sum_{\xx\in\T}
 \lrB{
 \max\lrb{\alpha_A,1}
 \sum_{V|U}
 2^{-n[D(\nu_{V|U}\|\mu_{Y|X}|\nu_U)
 +|R_A-H(U|V)|^+ -\lambda_{\X\Y}]}
 +
 \beta_A
 }
 \notag
 \\
 &
 \leq
 \frac {|\T|}{|\im\A||\im\B|}
 \lrB{
 \max\lrb{\alpha_A,1}
 2^{-n[F_{Y|X}(R_A)-2\lambda_{\X\Y}]}
 +
 \beta_A
 }
 \notag
 \\
 &\leq
 2\kappa
 \lrB{
 \max\lrb{\alpha_A,1}
 2^{-n[F_{Y|X}(R_A)-2\lambda_{\X\Y}]}
 +
 \beta_A
 },
 \label{eq:C2}
\end{align}
where
the second inequality comes from
Lemma \ref{lem:E} and (\ref{eq:C2sub}),
the third inequality comes from Lemmas~\ref{lem:exprob} and
\ref{lem:typenumber}, the fourth inequality comes from
the definition of $F_{Y|X}$ and Lemma~\ref{lem:typebound}
and the last inequality comes from  (\ref{eq:TU}).

From (\ref{eq:C0}), (\ref{eq:C1}),  and (\ref{eq:C2})
we have
\begin{align*}
 E_{ABC}\lrB{\Error_{Y|X}(A,B,\cc)}
 &\leq
 \alpha_{AB}-1
 +
 \frac{\beta_{AB}+1}{\kappa}
 +
 2\kappa
	\lrB{
 \max\lrb{\alpha_A,1}
 2^{-n[F_{Y|X}(R_A)-2\lambda_{\X\Y}]}
 +
 \beta_A
	}.
\end{align*}

Applying the above argument for all $\mu_{Y|X}$ satisfying
(\ref{eq:rateA}) and (\ref{eq:rateAB}),
we have the fact that there are $A\in\A$, $B\in\B$, and $\cc\in\im A$ 
that satisfy (\ref{eq:error-channel}).
\hfill\QED

\section{Conclusion}
The fixed rate universal coding theorems
are proved by using the notion of hash property.
We proved the theorems of fixed-rate lossless universal source coding 
and fixed-rate universal channel coding.
Since an ensemble of sparse matrices satisfies the hash property requirement,
it is proved that we can construct universal codes by using sparse matrices.

\appendix
We introduce the following lemmas that are used in the proofs of the theorems.

\begin{lem}[{\cite[Lemma 2.2]{CK}}]
 \label{lem:typebound}
 The number of different types of sequences in $\X^n$
 is fewer than $[n+1]^{|\X|}$.
 The number of conditional types of sequences $\X\times\Y$
 is fewer than $[n+1]^{|\X||\Y|}$.
 \hfill\QED
\end{lem}

\begin{lem}[{\cite[Lemma 2.3]{CK}}]
 \label{lem:typenumber}
 For a type $U$ of a sequence in $\X^n$,
 \begin{align*}
	2^{n[H(U)-\lambda_{\X}]}\leq |\T_U|\leq 2^{nH(U)},
 \end{align*}
 where $\lambda_{\X}$ is defined in (\ref{eq:lambda}).
 \hfill\QED
\end{lem}

\begin{lem}[{\cite[Lemma 2.6]{CK}}]
 \label{lem:exprob}
 \begin{align*}
	\frac 1n\log \frac 1{\mu_{X}(\xx)}
	&= H(\nu_{\xx})+D(\nu_{\xx}\|\mu_{X})
	\\
	\frac 1n\log\frac 1{\mu_{Y|X}(\yy|\xx)}
	&= H(\nu_{\yy|\xx}|\nu_{\xx})
	+D(\nu_{\yy|\xx}\|\mu_{Y|X}|\nu_{\yy}).
 \end{align*}
 \hfill\QED
\end{lem}

\begin{lem}[{\cite[Lemma 2]{CRYPTLDPC}}]
 \label{lem:typical-number}
 For $\yy\in\T_V$,
 \begin{align*}
	|\lrb{\xx': H(\xx')\leq H(U)}|
	&\leq
	2^{n[H(U)+\lambda_{\X}]}
	\\
	|\lrb{\xx': H(\xx'|\yy)\leq H(U|V)}|
	&\leq
	2^{n[H(U|V)+\lambda_{\X\Y}]},
 \end{align*}
 where $\lambda_{\X}$ and $\lambda_{\X\Y}$
 are defined by (\ref{eq:lambda}).
 \hfill\QED
\end{lem}
\begin{proof}
 The first inequality of this lemma is shown by the second inequality.
 The second inequality is shown by
 \begin{align*}
	|\lrb{\xx': H(\xx'|\yy)<H(U|V)}|
	&=
	\sum_{\substack{
	U':\\
	H(U'|V)\leq H(U|V)
	}}
	|\T_{U'|V}(\yy)|
	\\
	&\leq
	\sum_{\substack{
	U':\\
	H(U'|V)\leq H(U|V)
	}}
	2^{nH(U'|V)}
	\\
	&\leq
	\sum_{\substack{
	U':\\
	H(U'|V)\leq H(U|V)
	}}
	2^{nH(U|V)}
	\\
	&\leq
	[n+1]^{|\X||\Y|}
	2^{nH(U|V)}
	\\
	&=
	2^{n[H(U|V)+\lambda_{\U\V}]},
 \end{align*}
 where the first inequality comes from Lemma \ref{lem:typenumber}
 and the third inequality comes from
 Lemma \ref{lem:typebound}.
\end{proof}

\section*{Acknowledgements}
This paper was written while one of authors J.\ M.\ was a visiting
researcher at ETH, Z\"urich.
He wishes to thank Prof.\ Maurer for arranging for his stay.

\end{document}